\documentstyle[twocolumn,aps] {revtex}
\begin{document}
\title{ Manifestation of history dependent critical  currents  via
{\it dc} 
and {\it ac} magnetisation measurements in single crystals of
$CeRu_2$ and $2H-NbSe_2$\\
\bigskip
G. Ravikumar, V.C. Sahni, P.K. Mishra and T.V.
Chandrasekhar Rao\\
\it Technical Physics and Prototype Engineering Division,
Bhabha Atomic Research Centre, Mumbai 400 085, INDIA\\
S.S. Banerjee, A.K. Grover, S. Ramakrishnan and S.
Bhattacharya$^*$\\
\it Department of Condensed Matter Physics and Material
Science,
Tata  Institute  of  Fundamental  Research, Mumbai -400 005,
INDIA\\
M.J. Higgins\\
\it NEC Research Institute,4 Independence way, Princeton, NJ
08540\\
E. Yamamoto $^1$, Y. Haga$^1$, M. Hedo $^2$, Y.
Inada$^2$ and Y.Onuki$^{1,2}$\\
\it $^1$Faculty of Science, Osaka University, Toyonaka 560,
JAPAN\\
\it $^2$Advanced Science Research centre, Japan Atomic
Energy Research Institute, Tokai, Ibaraki 319-11, JAPAN\\
\bigskip
\small \vskip 0pt
\noindent \hfill
A  study of path dependent effects in single crystals of
$CeRu_2$
and {$2H-NbSe_2$} show  that  critical  current  density  $J_c$ 
of
the vortex  state  depends  on its thermomagnetic history over
a very
large part of $(H,T)$ parameter  space. The path dependence
in  $J_c$
is absent  above    the  peak position (i.e., $H  >  H_p$) of  the
peak effect region, which we believe identifies
the complete loss of order in the vortex structure. The highly
disordered FC  state can be healed into a relatively ordered
vortex
lattice by subjecting it to a large enough  change  in  {\it dc} 
field
(few  tens  of Oe) or by shaking the FC state with sufficient
{\it ac} field (few Oe).\\
\vskip 16pt
\noindent
\normalsize
Ms number: ~~~~~~PACS Numbers: 64.70 Pf, 74.60 Ge, 74.25
Ha, 74.70 Ad, 74.60 Jg}
\maketitle
\normalsize
\newpage
Investigating  structure  of  vortex lattice or flux line lattice
(FLL) in the mixed state of type II superconductors continues 
to
be   of   intense   interest.  Recent  theoretical  studies  have
postulated various glassy states in  FLL  arising  from  quenched
disorder and thermal fluctuations \cite{r1}. Experimental efforts
have  been  focused  on  detection  and  characterisation of such
states. The appearance of the Peak Effect (PE)  in  some 
systems
involving  an  anomalous  enhancement of critical current
density
$J_c$ in close proximity of the softening/melting in their FLL 
\cite{r2,r3,r4} -
has  been  explained  in  terms of a loss of spatial order in FLL
\cite{r5}. But the precise nature  of  this  loss  of
order  and its relationship to the glassy state of FLL are topics
of current debate.

In disordered magnetic systems such as  spin  glasses
\cite{r6}, one encounters the appearance   of  thermomagnetic 
history  effects. The temperature below which magnetisation
values under
zero  field  cooled (ZFC) and field cooled (FC) conditions differ
is usually identified as the spin  glass  transition  temperature
$T_g$ \cite{r6}. We find analogous manifestations in the
magnetic
behaviour   in   weakly  pinned  superconducting  systems,  viz.,
$2H-NbSe_2$(T$_c$=6.1K)   \cite{r7}   and   
$CeRu_2$(T$_c$=6.3K)
\cite{r8}, as reflected in their $J_c$ values. This difference in
the  FC  and  ZFC  response,  we  believe, reflects the different
extents of FLL correlations in  these  states,  and  is  seen  to
persist  upto the peak position of PE. In this sense the locus of
PE  in H-T plane may be regarded as the counterpart of spin
glass
transition temperature. The results of  {\it  dc}  and  {\it  ac}
magnetisation   experiments   in   the  two  superconductors  are
presented so as to elucidate history effects  well  before,  just
prior  to,  during  and  after  the  occurence  of peak of the PE
phenomenon. Our results {\it inter alia} add a newer facet to the
well known Critical State Model (CSM) \cite{r9} which 
postulates
a  unique critical current density $J_c$ for the vortex state for
a given field (H) and temperature (T).

We recall that hysteresis in magnetisation is related to $J_c(H)$
\cite{r10}. In isothermal magnetisation measurements, this
single
valued  $J_c$  translates  into a generic magnetisation hystersis
loop (see Fig.1) such that the {\it forward}  and  {\it  reverse}
branches   of   the   magnetisation   curve  define  an  envelope
\cite{r10}, within which lie all the  magnetisation  values  that
can be measured at the given temperature along various paths
with
different  thermomagnetic  histories  \cite{r11}.  For  instance,
Fig.1 schematically illustrates that the FC  magnetisation  curve
generated  by  decreasing the field after cooling the sample in a
{\it dc} field eventually merges into the  reverse  magnetisation
branch.  {\it  The  new  result of our experiments is that the
magnetisation curve, originating from a given FC
state, need not always be confined within the generic  hysteresis
loop}.  We  argue  that the observed new behaviour
elucidates the existence of  multivalued  nature  of  $J_c(H,T)$,
i.e.,  the  critical  current  density of vortex state at a given
$(H,T)$ depends on its thermomagnetic history. Our  inference 
by
an  equilibrium  {\it  dc}  magnetisation  technique strengthens
an 
earlier conclusion  from  non-equilibrium  transport  studies  on
$2H-NbSe_2$  by  Henderson et al \cite{r7}, who measured
that the
transport $J_c$ in ZFC state  is  considerably
lower  than  that  in  the  FC  state for fields below the peak
field
$H_p$.

{\it  AC}  susceptibility  measurements  in superimposed {\it
dc}
fields were performed on a  home  built  {\it  ac} 
susceptometer
\cite{r12}.  The  {\it  dc}  magnetisation measurements have
been
performed  using  Quantum  Design (QD) Inc.   (Model  
MPMS)   SQUID
magnetometer. The single crystals of cubic $CeRu_2$ and
hexagonal
$2H-NbSe_2$ were mounted on the sample holder such that the
field
is  parallel  to  cube edge and c-axis, respectively. Usually, the
measurement of magnetic moment $m$ in the MPMS SQUID
magnetometer
involves sample motion along the pickup coil array in the 
second
derivative  configuration,  over a scan length $2l$. The
magnetic
moment $m$ is obtained by fitting the  {\it sample  response} 
measured
over $-l < z < l$ to the form,

$$ V = a + bz + m c \phi (z-z_0),  \eqno(1)$$
where,
$$ \phi (z) = (\mu_0 R^2/2)~[-[R^2 + (z+Z)^2]^{-3/2}$$

$$ +2[R^2 + z^2]^{-3/2} - [R^2 + (z-Z)^2]^{-3/2}] \eqno(2)
$$

Here,  $a$,  $b$  and  $z_0$  account for constant offset, linear
drift and possible off-centering of the sample respectively.  $R$
(=  0.97  cm) is the radius and $2Z$ (= 3.038 cm) is the
distance
between the two outer turns of the pick-up coil array. $z$ is the
sample distance from the centre of the pickup coil array and 
$c$
is  the calibration factor. This analysis implicitly assumes that
$m$ is constant along the scan length and  therefore 
independent
of  $z$.  But, as described in Ref. \cite{r13}, when a
superconducting
sample, which exhibits PE, is moved in an inhomogeneous 
external
field,   its   magnetic   moment  can  become  strongly  position
dependent, leading to  spurious  experimental  artefacts  in  the
data.  Thus, an  appropriate method needs to be devised to
obtain
magnetisation values which are free from such artefacts. We
have
done this, by  analysing  the  raw  data  using  a  new  procedure
that  can  be termed as {\it half-scan} technique and
its salient features are detailed below.

In  5.5 Tesla QD MPMS model, on either side of the centre of
the magnet
\cite{r14} the field   due   to   the   superconducting    solenoid
monotonically  decreases  along  the axial direction. The central
idea of the {\it half scan} technique is  to  record  the {\it
sample
response} by  moving  it  over  that part of the axis so that the
sample does not experience field excursions. On the {\it
forward}
magnetisation curve, this is accomplished by recording the
sample
response  {\it  only}  between   $z=-l$   and   $z=0$.   As   the
magnetisation  of  the sample stays nearly constant for $-l < z
<
0$, we can fit this data to Eqn. 1 and obtain magnetic  moment
$m$ on
the  {\it forward} magnetisation curve. As illustrated in Fig.2a,
the SQUID response in the conventional measurement
(spanning $-l$
to $l$) fits very poorly to the ideal dipolar response  given  by
Eqn.1.  On  the  other hand the {\it half-scan} response
measured
between $z = -2$ $cm$ and $z =  0$  gives  an  excellent  fit  to
Eqn.1. The SQUID response of $2H-NbSe_2$ single crystal,
shown in
Fig.2  is  measured at 4.5K in a field of 8 kOe, i.e., very close
to peak field $H_p$. Similarly, to obtain the magnetic moment 
on  the  {\it
reverse}  magnetisation curve, the sample is initially positioned
at $z=0$ (i.e., where the field is maximum along the axis of  the
solenoid).  The  SQUID  response  shown  in Fig.2b is recorded
by
moving the sample between $z=0$ and  $z=l$.  Magnetic 
moment  is
then  obtained  by  fitting  this  response  to  Eqn.1. The SQUID
responses shown in Fig.2(a)  and  (b) have been compensated 
for  the
offset $a$ and the drift $bz$ (cf. Eqn. 1).

Figs.  3  and  4  summarize  the central results of magnetisation
hysteresis and {\it ac} susceptibility experiments in crystals of
$2H-NbSe_2$ ($2 \times 2 \times 0.4$ $mm^3$) and 
$CeRu_2$ ($3 \times 1.5 \times 0.8$ $mm^3$).  (It  may  be 
stated  here  that  the
$2H-NbSe_2$  crystal  is  from  the  same  batch  as  was used
by
Henderson et al \cite{r7} and $CeRu_2$ crystal is  the  one 
used
for  de-Haas  van Alphen studies earlier \cite{r15}. As
mentioned
earlier, both these superconducting systems are weakly pinned
and
the  crystal  pieces  chosen  for   present   measurements   have
comparable  levels  of quenched disorder in them \cite{r16}).
Figs.
3a and 4a display the magnetisation hysteresis loops in the PE
regime  of $2H-NbSe_2$  and  $CeRu_2$  respectively. The
pronounced increase in the hysteresis in the PE  region  of  both
$2H-NbSe_2$  and  $CeRu_2$  signify the anomalous increase
in the
critical current density at the onset of  PE.  Figs. 3a  and  4a,
also,   show the magnetisation curves measured in reducing
fields, after  having  cooled  the  samples  in  the
pre-selected  magnetic  fields to  a  given  temperature. The
pre-selected field cooled magnetisation states can be  identified
by  filled  diamonds lying on the dashed line in Figs. 3a and 4a.
Magnetisation of the FC sample  in  reducing  magnetic  field 
is
measured in the same way as the {\it reverse} magnetisation
curve
is  generated.  In {\it ac} susceptibility measurements, the PE
manifests via an enhanced (shielding)  diamagnetic  response.
Fig.3(b)  shows  the  plot  of  in-phase  {\it ac} susceptibility
($\chi^\prime$) vs H in ZFC and FC states in $2H-NbSe_2$ 
crystal
at 5.1~K and Fig. 4(b) shows similar results for $CeRu_2$
crystal
at  4.5~K.  The  $\chi  ^  \prime$  data points in FC states were
measured after cooling down the  sample  in  a  given  H  to  the
respective temperatures from the normal state.

It can be seen in Figs.3(a) and 4(a) that the magnetisation curve
measured  on  field  cooling in $H > H_p$ readily merges with
the
usual  {\it  reverse}  magnetisation  curve.  This  can  be  well
understood  within  the  framework of conventional Critical
State
Model\cite{r9,r10}, which assumes that $J_c$  is  single  valued 
function  of
(H,T)  (see Fig. 1). However, when field cooled in $H < H_p$,
the
magnetisation values obtained  by  reducing  the  external  field
initially {\bf overshoot} the  {\it reverse} magnetisation curve
(see
Fig. 3a  and  Fig. 4a).  On  further  reducing  the  field,   the
magnetisation  values  fall  sharply  and  FC magnetisation curve
merges into the  usual  {\it  reverse}  magnetisation  hysteresis
branch. The first observation that the magnetisation values
initially go beyond the conventional hysteresis loop is a clear
indication of $J_c$ at a given $H$ in the
FC  state ($J_c^{FC}$) being larger than that for the vortex
state at the same
$H$ value on the usual {\it reverse}  magnetisation  branch. 
The
later observation  that  the  FC  magnetisation curve eventually
merges
into the {\it reverse} magnetisation branch implies  that
the  FC  vortex state transforms to a more ordered ZFC like
state
as the vortex state adjusts to a large enough change ( $10^2$
Oe for $CeRu_2$ and $10$ Oe for 2H-NbSe$_2$)
in  the  external {\it dc} field. A neutron study \cite{r17} on a
crystal of $CeRu_2$ had shown that the FC state far  below
the PE region comprised much more finely divided blocks than
that
in  the  ZFC state. Keeping this in view, on the basis of present
results, it may be stated that the finely divided FC vortex state
heals to the more  ordered  ZFC  state  in  response  to  changes
induced by large external field variation.

The   {\it  ac}  susceptibility  data  in  Figs.  3(b)  and  4(b)
corroborate the above stated conclusions. As  per  a  CSM 
result \cite{r9},
$\chi  ^  \prime  =  -1  + \alpha h_{ac}/J_c$, where $\alpha$ is
a
shape and size dependent parameter and $h_{ac}$ is the  {\it 
ac}
field  amplitude, the higher diamagnetic response in the FC state
as compared to that for the ZFC state, reflects larger  $J_c$  in
the  former  state.  The  history  dependence  in $\chi ^ \prime$
response ceases above the peak position of the PE  region. 
Also,
at  very  low fields ($H < 1 kOe$), the difference between FC
and
ZFC $\chi ^ \prime$ response is seen to decrease, consistent
with
transport $J_c$ measurements of Henderson et al \cite{r7}.

In  the  Larkin-Ovchinikov  \cite{r18}  description of pinning in
superconductors, $J_c \propto V_c^{-1/2}$,  where  $V_c$  is 
the
volume  of  a  Larkin  domain  within  which  flux  lines  remain
correlated.  Smaller  $J_c$  in  the  ZFC   state   \cite{r7,r19},
therefore, corresponds  to a more ordered FLL than in the case
of
FC state. For $H > H_p$, flux  lines  form a  quasi-pinned
state,  which  appears  to  be  independent  of  how the state is
approached in (H,T) space. For $H < H_p$, the  larger 
$J_c^{FC}$
can  therefore  be  attributed  to the formation of a more finely
divided disordered state, with concomitant  more  pinning. 
While
subjecting the FC state to a decrease in the external field, this
state  eventually  goes  over  into a relatively more ordered ZFC
state with a larger $V_c$, as manifested by a steep  fall  in  the
magnetisation   values  (after  overshooting  the  {\it  reverse}
magnetisation branch). A change from a disordered state  to 
more
ordered  vortex state can also be brought about by other kinds
of
perturbations as well. For example, in our {\it ac}
measurements,
we observed that the large (shielding) diamagnetism of  FC 
state
suddenly  collapses  to  that  of the ZFC state on increasing the
{\it ac} field amplitude momentarily to about 5 Oe.  Although 
by
no means obvious this way of {\it annealing} away the disorder
of
the FLL in the FC case is akin to annealing by a passage of large
transport current \cite{r7}.

$CeRu_2$  and  $2H-NbSe_2$  are  very  dissimilar
superconducting
systems as regards their microscopic physics;  the  former  is  a
mixed  valent system whereas the latter is a layered
chalcogenide
which exhibits charge density wave behaviour in its normal
state.
In the  context  of  vortex  state  of  superconductors,  the  PE
phenomena  in weakly pinned samples \cite{r2,r7,r8} of these
two systems
has been in current focus (apparently) due to different  reasons.
PE   in  $CeRu_2$  has  (often)  been  considered  to  relate  to
realization of Generalized Fulde-Ferrel Larkin Ovchinikov
(GFFLO)
state \cite{r8}, whereas in very clean samples of $2H-NbSe_2$,
PE
is ascribed  to  FLL  softening  \cite{r2}.  Since  normal  state
paramagnetism  of $2H-NbSe_2$ is small, it could not be a
serious
candidate for the occurence of GFFLO  state.  The  present 
findings,
that  the  experimental  features of the mixed state prior to and
across the PE  region  of  these  two  systems  follow  identical
course,  would  support  the  view that their behaviour in the PE
regime presumably reflects same generic physical  phenomenon
that occurs in the  
mixed  state in a weakly pinned flux line lattice
while approaching $H_{c2}$.

To  conclude,  we have demonstrated through {\it dc} and {\it
ac}
magnetisation measurements with a new {\it half-scan}
technique, in the mixed state  of  $CeRu_2$  and
$2H-NbSe_2$, that there are sizable thermomagnetic  history 
effects  in  their  critical  currents below  $H_p$, where  the 
peak  of the PE occurs.  We have shown that these
effects imply a more finely divided disordered vortex
arrangement
for the FC state, as compared to  that  for  the  ZFC  state.  It
should  be  noted that the critical current, remains finite above
$H_p$. This suggets that the glassy state above $H_p$  is 
weakly
pinned  and a change to completely unpinned state does not
appear
until the  higher  field  $H_{irr}$.  The  implication  of  these
results  with  respect to the occurence of PE behaviour,
fishtail (second peak), etc., in the Cuprate  superconductors, 
such
as,  YBCO  and  BSCCO,  remains  an  interesting topic for
further
investigations.

$^*$Present and permanent address: NEC Research Institute,
4 Independence Way, Princeton, NJ 08540, USA.

\begin{figure}
\caption{A schematic showing
magnetisation hysteresis curves in an irreversible type II
superconductor.
{\it Forward} and {\it reverse} magnetisation branches
corresponding to
increasing and decreasing  field cycles. A magnetisation curve
measured during reducing field cycle after cooling in a field is
indicated as field cooled (FC) curve. $M_{FC}$ denotes the FC
magnetisation value.}
\label {Fig.1}
\end{figure}
\begin{figure}
\caption{
(a)($\circ$) SQUID response of the sample in the PE region, on
the {\it forward} magnetisation curve using a conventional full
symmetric scan
of 4cm length. The  corresponding fit to Eqn.1 is shown by a
dotted line. The {\it half-scan} response for $-2 cm < z < 0$
($\bullet$) and its fit to Eqn.1 (continuous line) is also shown.
(b) The SQUID response from 0 to 2 cm  for the {\it reverse}
case is shown along with the corresponding fit to Eqn.1 after
compensating for
the  offset $a$ and linear drift $bz$.}
\label {Fig.2}
\end{figure}
\begin{figure}
\caption{
(a) A portion of the magnetisation hysteresis curve
(encompassing the PE region) recorded at
5.1K for $H \parallel c$ using the {\it half-scan} technique in a
$2H-NbSe_2$ crystal. Also shown are the magnetisation values
recorded while decreasing the field after cooling  the sample in
(pre-selected) different external
fields. The initial $M_{FC}$ values are identified by filled
diamonds lying on the dashed curve. Each FC magnetisation
curve initiates from a different $M_{FC}$ value. (b) {\it AC}
susceptibility measured with $h_{ac}$ $=$ 0.5 Oe at $f$ = 211
Hz, for $H \parallel c$ at 5.1K  (i) after cooling the sample 
in  zero field (ZFC) and (ii) after cooling the sample different
fields each time (FC). $\chi^ \prime (H)$
values are normalized to $\chi^ \prime (0)$ }
\label {Fig.3}
\end{figure}
\begin{figure}
\caption{
(a) A portion of the magnetisation hysteresis curve of
$CeRu_2$ recorded at 4.5K for $H \parallel$ [100] using the
{\it half-scan} technique. The FC magnetisation curves
originating from different $M_{FC}$  values are also shown.
(b) {\it AC} susceptibility for $H \parallel$ [100] in both ZFC
and FC modes as described in the caption of Fig.3b.}
\label {Fig.4}
\end{figure}

\begin{references}
\bibitem{r1} T. Giamarchi and P. Le Doussal, Phys. Rev. Lett.
{\bf 72}, 1530 (1994);
Phys. Rev. B {\bf 52}, 1242 (1995); M. Gingras  and  D.A.
Huse,
Phys. Rev. B {\bf 53}, 15193 (1996); G. Blatter et al, Rev.
Mod. Phys.
{\bf 66}, 1125 (1994) and references therein.
\bibitem{r2} S. Bhattacharya and M.J. Higgins, Phys. Rev. Lett.
{\bf 70}, 2617 (1993); Phys. Rev. B {\bf 52}, 64 (1995); 
Physica C  {\bf 257}, 232 (1996)
and references therein.
\bibitem{r3} W.K. Kwok et al, Phys. Rev. Lett. {\bf 73}, 2614
(1994)
\bibitem{r4} A.I. Larkin et al, Phys.Rev.Lett. 75, 2992 (1995);
K. Ghosh et al, ibid, {\bf 76}, 4600 (1996).
\bibitem{r5} T.V.C. Rao et al, Physica C (in press); S.S.
Banerjee et al (unpublished).
\bibitem{r6} J.A. Mydosh, Spin glass: An experimental
introduction, Taylor and Francis, London (1993)
\bibitem{r7} W. Henderson et al, Phys. Rev. Lett. {\bf 77}
2077 (1996).
\bibitem{r8} R. Modler  et al, Phys. Rev. Lett. {\bf 76}, 1292
(1996); Czech J. Phys. {\bf 46}, Suppl. S6, 3123 (1996); A.
Yamashita et al, Phys. Rev. Lett. {\bf 79}, 3771 (1997).
\bibitem{r9} C.P. Bean, Rev. Mod. Phys. {\bf 36}, 31 (1964).
\bibitem{r10}P. Chaddah, K.V. Bhagwat and G. Ravikumar,
Physica C {\bf 159}, 570 (1989) and references therein.
\bibitem{r11} A.K. Grover et al, Physica C {\bf 162-164}, 337
(1989);
Pramana-J Phys {\bf 33}, 297 (1989); B.V.B. Sarkissian et al,
ibid {\bf 38}, 641 (1989).
\bibitem{r12} S. Ramakrishnan et al, J. Phys. E {\bf 18}, 650
(1985).
\bibitem{r13} G. Ravikumar et al, Physica C {\bf 276}, 9
(1997).
\bibitem{r14} Qunatum Design Technical Advisory MPMS
No.1
\bibitem{r15} M. Hedo et al, J. Phys. Soc. of Japan {\bf 64},
4535 (1995).
\bibitem{r16} Transport $J_c$ values in $CeRu_2$ and
$2H-NbSe_2$ are estimated to be $\sim$ $10^2~~A/cm^2$ and
$\sim$ $10^3~~A/cm^2$, respectively.
\bibitem{r17}A.D. Huxley et al, Physica B {\bf 223-224}, 169
(1996).
\bibitem{r18} A.I. Larkin and Yu.N. Ovchinikov, J.Low Temp.
Phys. {\bf 34}, 409 (1979).
\bibitem{r19}  N.R. Dilley et al, Phys. Rev.B {\bf 56}, 2379
(1997).
\end{references}
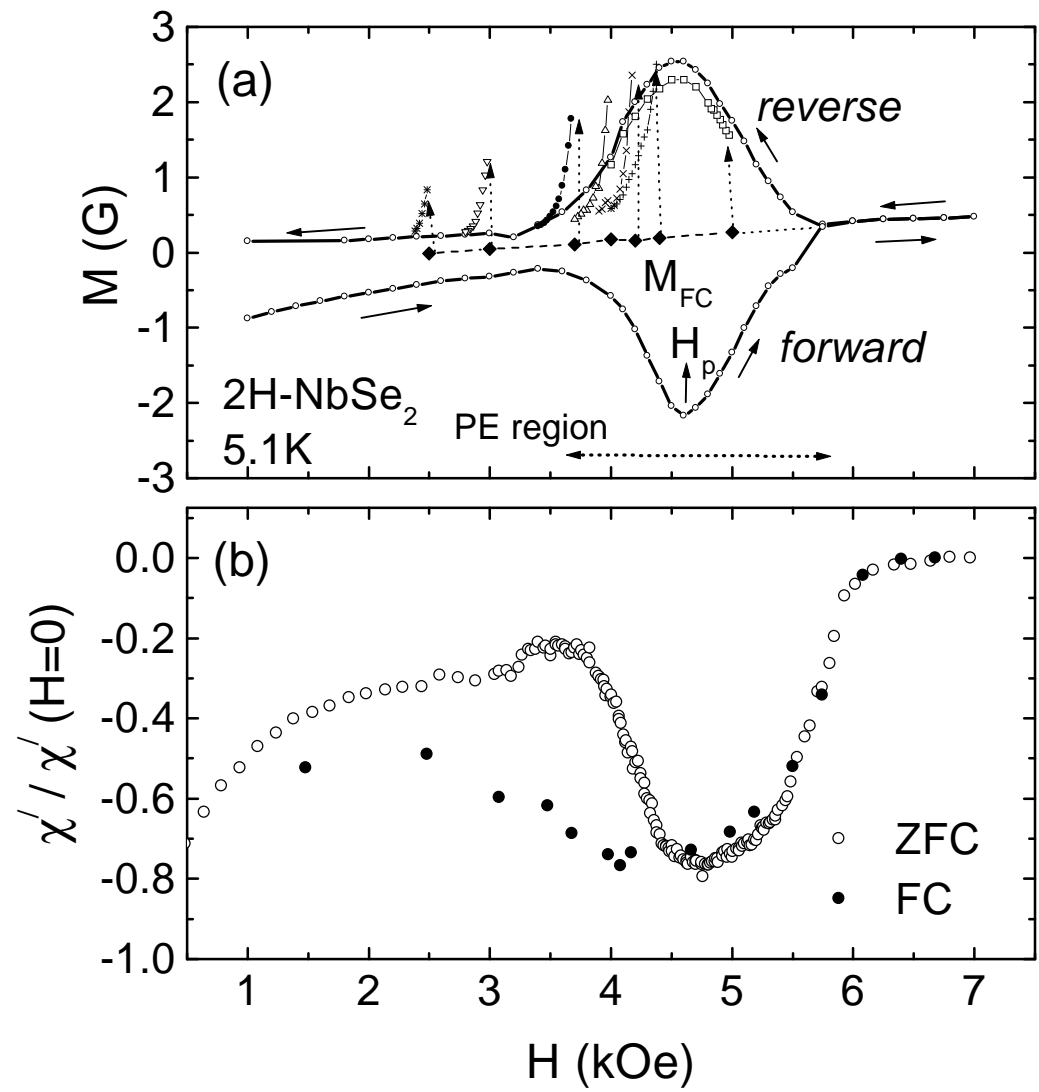
\end{document}